\documentclass[usenatbib,useAMS]{mn2e}

\usepackage{epsfig}
\usepackage{subfigure}
\usepackage{amsmath}
\usepackage{amssymb}

\begin{document}

\title[The gravitational wave signal from close double degenerates]
{The influence of star formation history on the gravitational wave signal from close double degenerates in the thin disc}

\author[S.\,Yu \& C.S.\,Jeffery]
   {Shenghua Yu and C. Simon Jeffery
   \\
   Armagh Observatory, College Hill, Armagh BT61 9DG,
   N. Ireland
}

     \date{Accepted .
           Received ; }

      \pagerange{\pageref{firstpage}--\pageref{lastpage}}

\label{firstpage}

    \maketitle

\begin{abstract}

The expected gravitational wave (GW) signal due to double degenerates
(DDs) in the thin Galactic disc is calculated using a Monte Carlo simulation.
The number of young close DDs that will  contribute observable discrete
signals in the frequency range $1.58 - 15.8$ mHz is estimated by comparison
with the sensitivity of proposed GW observatories. The present-day DD population
is examined as a function of Galactic star-formation history alone. It is shown that
the frequency distribution, in particular,  is a sensitive
function of the Galactic star formation history and could be used to measure the
time since the last major star-formation epoch.
\end{abstract}


\section{Introduction}
\label{sec_introduction}

Theoretical studies  have indicated that double
degenerate (DD) stars with very close orbital separation
should be the most abundant and prominent
sources of gravitational wave (GW) radiation at frequencies between
0.1 and 30 mHz \citep{Peters63,Landau75,Evans87}. In addition to
other binaries, they will form a GW confusion background due to the
large number of sources per frequency resolution bin
\citep{Hils90,Hils00}.

In general, the number of DDs per equal-interval frequency bin
at high frequency in any Galactic population drops as a function of frequency.
For a frequency resolution better than $\sim3\times10^{-8}$ Hz and frequencies
$\gtrsim$1 mHz (i.e. orbital periods $\leq 33$min), a GW experiment may detect
only one or no individual sources per resolution bin, making the source potentially
resolvable\footnote{If there is only one DD in a resolution bin and
the GW signal from this DD is greater than a given detection threshold
(e.g. S/N=3 in this paper), we define the DD to be a resolved source. If the GW signal
from this DD is less than the same detection threshold
(e.g. S/N=3 in this paper), we define the DD to be a potentially resolved source.} \citep{Evans87,Hils00,Nelemans01b,Ruiter09,Yu10}.
Resolved DDs identified from GW detections would allow us to study
the relation between their properties and the dynamic nature of
globular clusters \citep{Willems07}, the nature and potential outcome
of the DD merging process \citep{Webbink84}, and their association with
the formation and evolution of a galaxy.

Since DDs in a galaxy are the outcome of stellar formation and evolution, a key
question is how the star formation (SF) history, the distribution of initial orbital
parameters and the evolution of their main-sequence progenitors influence
the DD population in terms of final distribution with orbital
period (GW frequency), separation and chirp mass\footnote{Chirp mass ($\mathcal{M}$)
of a binary is defined as $\mathcal{M}\equiv\left(\frac{m_{1}m_{2}}{m_{1}+m_{2}}\right)^{3/5}\left(m_{1}+m_{2}\right)^{2/5}$
where $m_{1}$ and $m_{2}$ is the mass of each component of the binary respectively. A general
relation between the GW frequency $f$ and the orbital period $P_{\rm orb}$ in the $n$th harmonic
is $f=\frac{n}{P_{\rm orb}}$. In particular, if a binary is in a circular orbit (eccentricity $e$=0),
we have $f=\frac{2}{P_{\rm orb}}$.}.
This question can be addressed using the method of binary-star population synthesis (BSPS)
to investigate, for example, the SF history alone.  In our previous papers \citep{Yu10,Yu11},
the influence of star formation and evolution on the DDs in a galaxy was characterised by the
use of a parameteric SF rate and synthetic SF contribution functions.

The SF history in a galaxy is likely to be more complicated than
approximations used in previous studies, for example, a single star-burst, constant
SF, or (quasi-) exponential SF  (\citet{Nelemans01b,Ruiter09,Yu10}).
Observations agree that the SF rate in the Galactic disc is likely to
be declining but is episodically enhanced \citep{Majewski93,Rocha-Pinto00}.
The aim of this paper is to show how the GW signal from the DD
population in the thin disc in the Galaxy depends on the SF history whilst
keeping other factors in the model constant.

We compute the GW signal from a simulated DD population
in the thin disc of the Galaxy in \S\,\ref{sec_simulation}, and compare
the results for three different SF models in
\S\,\ref{sec_comparison}.
We compare our results with previous studies in \S\,\ref{sec_discussion},
and draw significant conclusions in \S\,\ref{sec_conclusion}.

\section{The GW signal from resolved double degenerates in the thin disc}
\label{sec_simulation}

We simulate the DD population in the thin disc of the Galaxy by
adopting a simplified binary-star evolution and population-synthesis model (BSPS).
Inputs include a star formation rate (SFR),
an initial mass function (IMF), a mass ratio distribution $P_q$, an orbital separation distribution $P_a$,
an eccentricity distribution $P_e$, a metallicity $Z$ and a representation of the thin disc structure \citep{Yu10}.
If the orbital period of a binary is sufficiently long (i.e. $\gtrsim$100 yrs), the binary
is assumed to behave as two single stars and will not form a DD. Inputs also include binary-star evolution physics,
including mass-loss rates, common-envelope ejection efficiency, etc.,  described by \citet{Yu10}.

To provide a benchmark calculation, the SFR is approximated by a quasi-exponential function
\begin{equation}
S(t)=7.92 e^{-t/\tau}+0.09t~~{\rm    M_{\odot}yr^{\rm -1}}
\end{equation}
where $t$ is the age of the disc and $\tau=9$ Gyr, yielding an SFR $\approx 3.5 {\rm M}_{\odot}$yr$^{-1}$
and a total disc mass (in stars)
$M_{\rm d}\approx5.2\times10^{10} {\rm M}_{\odot}$ at the current epoch ($t=10$ Gyr).
The value for the current SFR is consistent with recent observations \citep{Diehl06}.
We adopt the IMF of \citet{Kroupa93},
a constant $P_q$ \citep{Mazeh92,Goldberg94}, and
$P_a$ following \citet{Han98} and \citet{Griffin85}.
$P_e=2e$  follows \citet{Nelemans01b}.
We adopt $Z=0.02$ (solar) for all thin disc binaries.

The simulation is started with a sample of $10^{7}$ primordial MS+MS binaries, with total mass
$\approx1.05\times10^{7}~{\rm M}_{\odot}$. Initial parameters for each binary are allocated on
the above distributions using a Monte Carlo procedure.  This procedure yields
a total of $\sim4.93\times10^{4}$ DDs over a period of 15 Gyr. We compute
the birth rate, merger rate and total number, and record their individual properties
(masses and orbital parameters) as a function  of  the age of the thin disc \citep{Yu11}.

Subsequently, a second Monte Carlo procedure is used to generate the orbital properties
for a complete sample of $n_{\rm dd}$ thin-disc DDs by interpolation on the initial DD
sample described above. The initial DD sample obtained from computations of stellar evolution
and the complete sample are identical with \citet{Yu11}, and similar to \citet{Yu10}
(the difference to the total number DDs in the thin disc is $\sim8\%$).
We identify four types of DD pair by the chemical composition of their degenerate cores,
{\it i.e.} He+He, CO+He, CO+CO, and ONeMg+X where X = He, CO, or ONeMg. Since we use a
simplified model, the criteria for identifying the WD core compositions are given by the
initial-final mass relation \citep{Han98,Hurley00,Hurley02}.

\citet{Yu11} give $n_{\rm dd}$ for the four different types of DD in this simulation.
These $n_{\rm dd}$ DDs are distributed in a thin disc model where the
stellar mass density is described as
\begin{equation}
\rho_{\rm d}(R,z) = \frac{M_{\rm d}}{4\pi h_{R}^{2}h_{z}} e^{-R/h_{R}}\textrm{sech}^{2}(-z/h_{z})
 \quad {\rm M_{\odot} pc^{-3}},
\end{equation}
and where $R$ and $z$ are the natural cylindrical coordinates of the
axisymmetric disc, and $h_{R}=2.5$\,kpc  and $h_{z}=0.352$\,kpc
are the radial scale length and scale height of the thin disc \citep{Sackett97,Phleps00}.
This structure and the thin disc mass (in stars)
are consistent with the dynamical parameters given by \citet{Klypin02} and \citet{Robin03}.

The DDs are sorted to obtain the number distribution by orbital frequency, to calculate the
total strain amplitude $\sum h^{2}$ from all DDs in each frequency bin, and hence
to obtain the GW strain amplitude-frequency relation due to the thin disc DDs.
We adopt a frequency resolution element $\Delta f = 3.17\times10^{-8}\,{\rm  Hz}$,
corresponding to one year of observation with a GW detector.

\begin{figure*}
\centering
\includegraphics[width=18cm,clip]{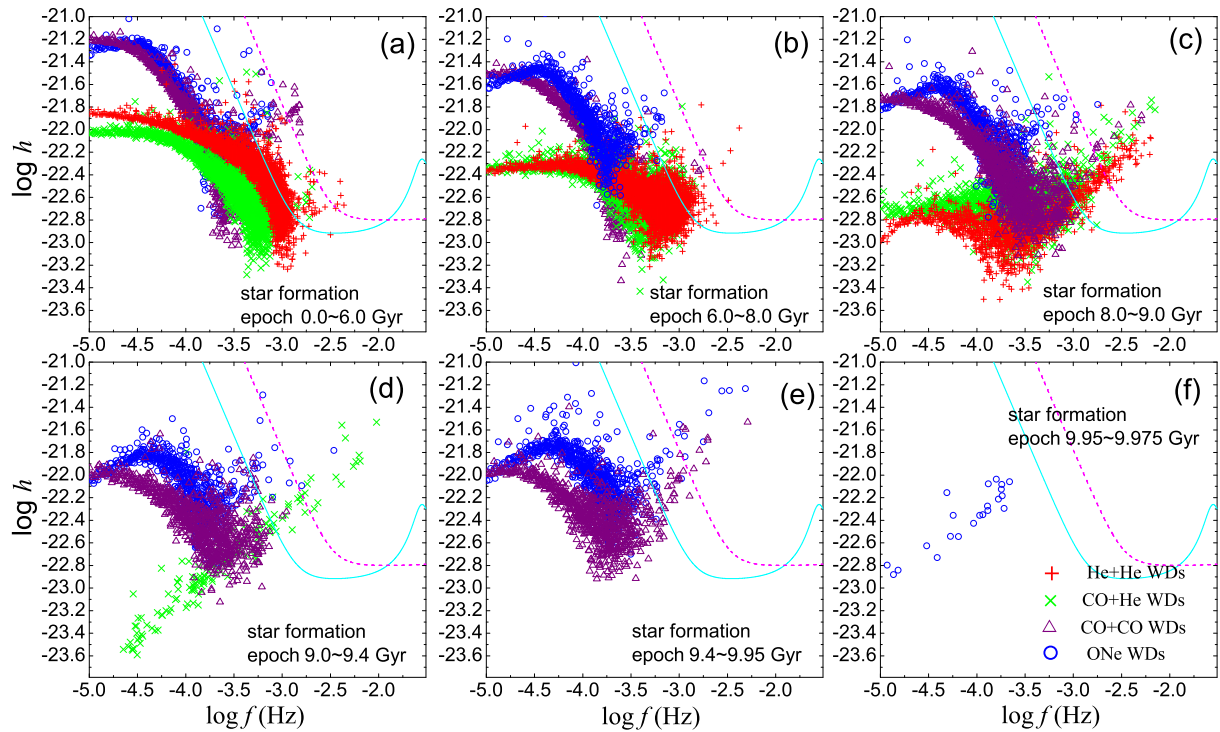}
\caption{The contribution of the current DD populations from different star-formation epochs to the
gravitational-wave amplitude ($\log h$) in the quasi-exponential star-formation model.
Solid (cyan) and broken (magenta) lines represent the eLISA and LAGRANGE sensitivity with S/N=3 for a one year mission,
respectively.}
\label{fig_sfgws}
\end{figure*}

Figure \ref{fig_sfgws} illustrates the influence from different SF epochs
on the predicted GW signal. The expected sensitivities for S/N=3 for
the proposed GW detectors eLISA and LAGRANGE (or eLISA-type mission) in
a one year mission are also shown  \citep{Larson00,Conklin11}.

From this figure, we identify a turning point, or critical frequency $f_c$,
around 0.001 Hz. For frequencies $f<f_{\rm c}$, the amplitude of the GW varies approximately
{\it inversely} with frequency, forming a `continuum'. For frequencies above the
critical frequency $f_c$, the amplitude becomes a discrete signal.

The existence of the critical frequency
is strongly associated with the product of the number distribution of DDs with respect
to the frequency and the size of the frequency element, {\it i.e.} $\frac{{\rm d} n(f)}{{\rm d} f}\Delta f$.
When $f\leqslant f_{\rm c}$, i.e. $\frac{{\rm d} n(f)}{{\rm d} f}
\Delta f \geqslant 1$ , there must be one or more DDs per resolution
element $\Delta f$, implying that $n(f)$ can be represented by a continuous
function.
When $f>f_{\rm c}$, {\it i.e.} $\frac{{\rm d} n(f)}{{\rm d} f}\Delta f < 1$ ,
some frequency bins will not contain any DD. Rather, there will be one
DD in $i=(\frac{{\rm d} n(f)}{{\rm d} f}\Delta f)^{-1}$ frequency bins, i.e.
in one frequency bin between $f$ and $f+i \Delta f$.

These results are consistent with an analytic approach (e.g. \citet{Evans87}),
whereby
\begin{equation}
f_{\rm c} = 2.977
\left(\frac{\mathcal{M}}{M_{\odot}}\right)^{-5/11}
\left(\frac{\nu}{{\rm yr^{-1}}}\right)^{3/11}
\left(\frac{\Delta f}{10^{-8}}\right)^{3/11} {\rm mHz},
\label{eq_resolvedf}
\end{equation}
in which the chirp mass $\mathcal{M}$ of a binary is assumed to be an independent
variable and $\nu \equiv \int\nu(f){\rm d}f$ is the total birth rate
of DDs.

\begin{figure*}
\centering
\includegraphics[width=12cm,clip]{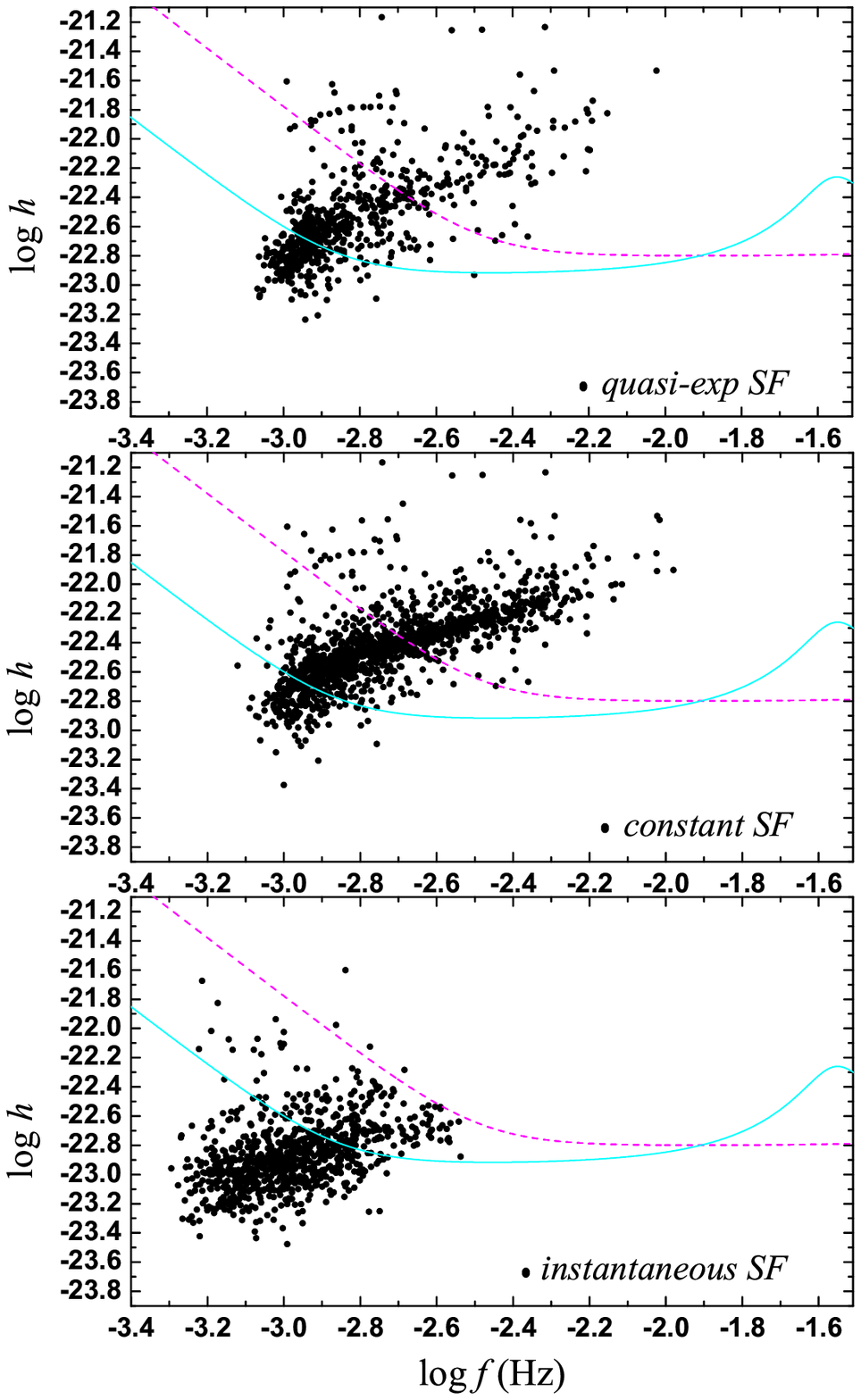}
\caption{Resolved and potentially resolved DDs in the different star-formation models. From top to bottom,
the SF models are quasi-exponential, constant and instantaneous (\S\,\ref{sec_comparison}).
Solid (cyan) and broken (magenta) lines represent the eLISA and LAGRANGE sensitivity with S/N=3 for a one year mission, respectively. For clarity, the systems shown represent a small sample (one DD in twenty)
of the complete thin-disc simulations.}
\label{fig_resolved}
\end{figure*}

Figure\,\ref{fig_sfgws} shows that the continuum of the thin disc GW
signal consists mainly of signal from early SF (i.e. $0.0-6.0$ Gyr).
Late SF  (i.e. 8.0$-$9.95 Gyr) contributes most of the discrete signal in the high
frequency band and only a small fraction of the continuum.
Sources with $\log f > -2.3$ principally comprise CO+He and He+He DDs
formed during the last epoch of SF. In the range $-3 < \log f <-2.3$,
different types of DD will be mixed, with CO+CO and ONeMg+X DDs dominating
the high amplitude signal and CO+He and He+He DDs contributing to the low
amplitude. We are primarily interested in these resolved sources since
their strain amplitudes exceed the sensitivities of the proposed  GW
detectors.

In addition, distance strongly affects the GW signal, both continuum
and discrete. The structure of the thin disc implies that the
distance to the majority of DDs will be large, but there will be a few
nearby (local) DDs. Thus, in our simulations, there are few isolated
strong signals due to DDs at $d<1$\,kpc
({\it e.g.} Fig.\,\ref{fig_sfgws}(b): $\log f=-2.88$ and $\log f=-2.98$).

\section{Comparison of different star formation models}
\label{sec_comparison}

We have discussed the GW signal with a {\it `Quasi-Exponential
   SF'} rate.  In order to see the influence on the GW signal
from different SF models, we have calculated the GW signal assuming
two further models for the SF rate. These are
{\it `Instantaneous SF'}, in which a single star burst takes place
only at the formation of the thin disc with a constant SF rate of
132.9$M_{\odot}$ yr$^{-1}$ from 0 to 391 Myr, and
{\it `Constant SF'}, in which SF occurs at a constant rate of
5.2$M_{\odot}$ yr$^{-1}$ from 0 to 10 Gyr.
All three models produce a thin-disc star mass of $\sim5.2\times10^{10}$
$M_{\odot}$ at the thin-disc age $t=10$ Gyr.

The differences between the signals in these models
arise from DDs with different progenitor properties. According to binary-star
evolution theory, main sequence stars with mass $M \leq 2 {\rm M}_{\odot}$ can
leave low-mass white dwarfs (WDs) with core mass $\lesssim$0.5 ${\rm M}_{\odot}$
if there is no significant mass accretion (e.g. \citet{Hurley02}).
Stars with $M\approx 2 - 10{\rm M}_{\odot}$ can develop WDs with massive
degenerate cores ($\gtrsim$0.5 ${\rm M}_{\odot}$) and heavier elements
({\it i.e.} CO or ONeMg). Since the evolution time-scale of a star is a strong
function of its initial mass, the birth rates of different WD types and the GW
signal are sensitive to different SF epochs. For example, the signals at high
frequency from He+He and CO+He DDs are sensitive to SF between $6-9$ and $8-9.4$
Gyr respectively (Fig.\,\ref{fig_sfgws}). Our results show that a constant SF
model produces a distinctly higher strain amplitude from He+He DDs ($-22.2<\log
h<-23.0$, $-2.8<\log f<-2.0$) than both other models. The quasi-exponential model
produces less high-frequency GW radiation than the constant SF model.

However, a very different discrete GW signal at $f>0.001$ Hz is
produced by the {\it instantaneous SF} model. The bottom panel in
Fig.\,\ref{fig_resolved} shows only a few GW signals from close DDs
at $f\gtrsim0.00158$ Hz ($\log f\gtrsim -2.8$). Since the majority of
close DDs merged in the past 10 Gyr
(see birth and merger rates in \citet{Yu11}), only a few close DDs now remain.
This implies that short-period DDs producing the discrete GW signal
are sensitive to recent SF (see Fig.\,\ref{fig_sfgws}).

\begin{table*}
%\begin{minipage}[t]{\columnwidth}
\caption{Proportion of each type of resolved DD in each SF model.}
\label{tab_resolved}
\begin{center}
\begin{tabular}{lccccccc}
 \hline
         & quasi-exp           & constant         & instantaneous \\
 \hline
 types of DDs\\
 He+He   & 71.0\%  & 78.9\% & 74.6\% \\
 CO+He   & 22.6\%  & 16.2\% & 17.1\% \\
 CO+CO   & 5.5\%   & 4.1\%  & 7.8\% \\
 ONeMg+X & 0.9\%   & 0.8\%  & 0.4\%  \\
 \hline
 total   & 11880   &22920   & 16240   \\
 \hline
 resolved by LAGRANGE with S/N=3   & 2690  & 9020   & 80   \\
 resolved by eLISA with S/N=3   & 8010  & 19820   & 3840   \\
 \hline
\end{tabular}
\end{center}
%\end{minipage}
\end{table*}

To represent possible future observations,
the sensitivities of the proposed eLISA and LAGRANGE missions are
shown in each figure.
In our present simulations, the number of (one-year) resolvable DDs are 11880,
22920 and 16240 for the quasi-exponential, constant, and instantaneous
SF models respectively, in which about 2690, 9020 and 80 could be resolved
by LAGRANGE with S/N=3, and about 8010, 19820, 3840 could be resolved by eLISA with
S/N=3. These systems are shown in Fig.\,\ref{fig_resolved}.
The proportion of each type of DD is shown in Table\,\ref{tab_resolved}.
The proportions of resolvable DDs detectable by LAGRANGE are quite different from those given
by the models because of the frequency response of the detector.
LAGRANGE is more sensitive to high-$f$ DDs, of which the highest proportion are found in the
constant SF model, which contains the highest fraction of DDs from recent star formation.

\begin{figure*}
\centering
\hspace*{-4cm}
\includegraphics[width=23cm,clip]{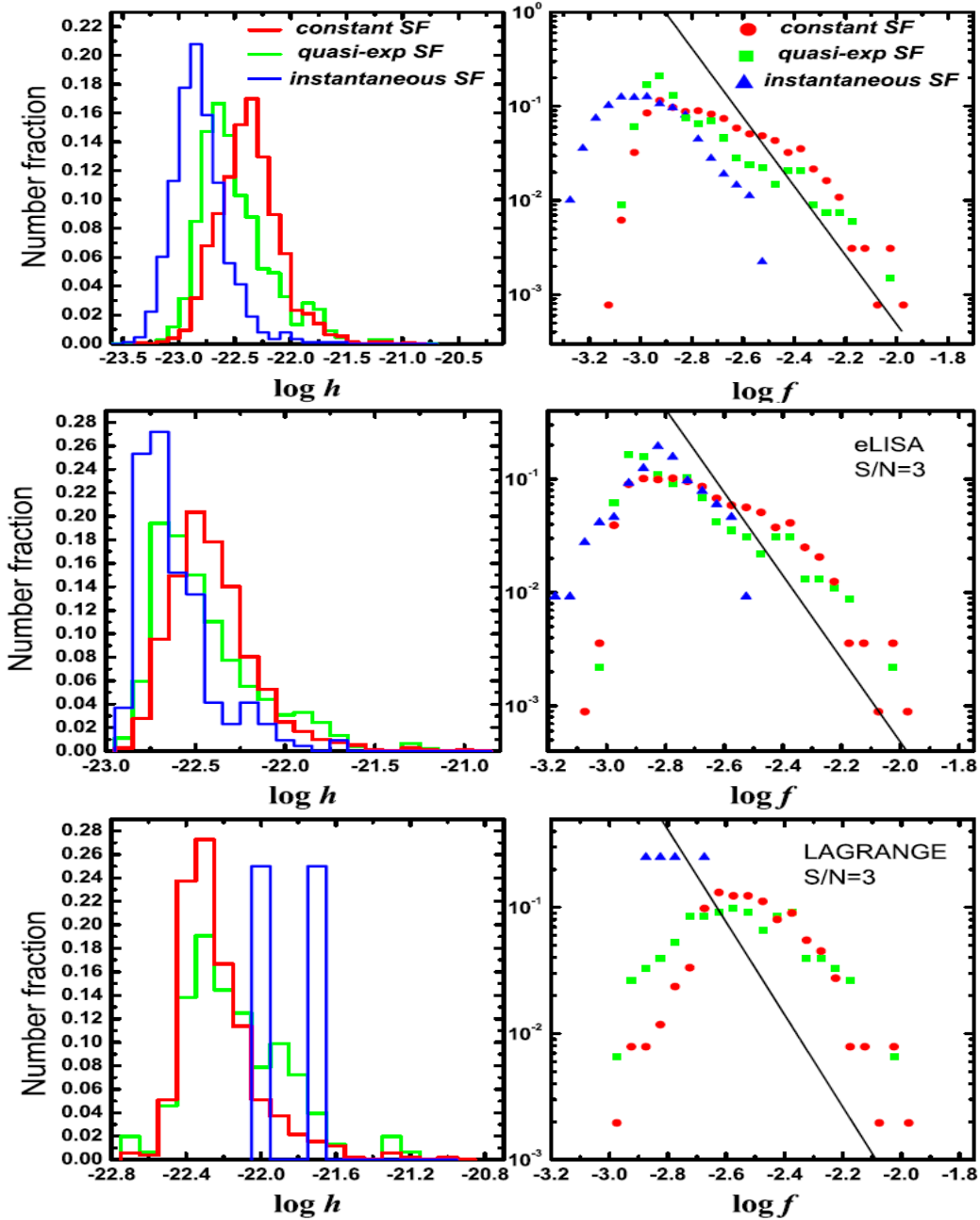}
\caption{Histogram of the GW signal distribution from resolvable DDs, including resolved
and potentially resolved DDs (top panels) and from resolved DDs by the two GW detectors (middle and
bottom panels) with respect to strain amplitude (left ) and frequency (right ) in different SF models, with
$\Delta \log h = 0.1$ and
$\Delta \log f = 0.05$. Blue represents the instantaneous star formation (SF) model,
red, constant SF, and green, quasi-exponential SF. The straight
black line represents the number-frequency relation: $\frac{{\rm d} n(f)}{{\rm d} f}\approx 0.0007 f^{-11/3},~
{\rm d}f=1~{\rm yr}^{-1}$.
Top panels: no correction for detector sensitivity, Middle: relative distributions assuming an eLISA-type detector with S/N=3,
Bottom: relative distributions assuming a LAGRANGE-type detector with S/N=3.
All number fractions are normalized to the total numbers of resolved DDs in each model (Table 1).  }
\label{fig_logh}
\end{figure*}

Figure\,\ref{fig_logh} shows the distribution of the GW strain amplitude
and frequency from all resolvable DDs in each SF model.
This figure can be understood analytically.

The change of orbital separation $\dot{a}$ of a DD in a circular orbit
with respect to time due to GW radiation is \citep{Landau75}
\begin{equation}
\dot{a} = -\frac{64}{5} G^{3}c^{-5}\mu M^{2}a^{-3}.
\label{eq_dadt}
\end{equation}
where $a$ is the orbital separation; $G$ is the gravitational
constant; $c$ is the speed of light; $\mu$ and $M$ are the reduced
mass and total mass of the DD, respectively. Combining
Eq.\,\ref{eq_dadt} with Kepler's third law $P_{\rm
  orb}^{2}/a^{3}=4\pi^{2}/GM$ and the relation between orbital period
and GW frequency $P_{\rm orb}=2/f$, we have the change of the GW
frequency of a DD per unit time
\begin{equation}
 \dot{f} = -\frac{96}{5} G^{5/3}c^{-5}\pi^{8/3}\mathcal{M}^{5/3}~f^{11/3},
\end{equation}
where $\mathcal{M}\equiv\mu^{3/5}M^{2/5}$ is the chirp mass (see footnote 2).
The inverse $\dot{f}^{-1}$ indicates the GW radiation timescale or
lifetime.

Assuming that the birth rate of the DDs at frequency $f $ is $\nu(f)$,
the number of DDs per unit frequency may be written as the product of birth rate and lifetime;
using appropriate units
\begin{equation}
\begin{split}
\frac{{\rm d} n(f)}{{\rm d} f} &= \nu(f)\cdot\dot{f}^{-1} \\
& = \frac{\nu(f)}{ \frac{96}{5} G^{5/3}c^{-5}\pi^{8/3}\mathcal{M}^{5/3}~f^{11/3}} \\
& \approx 0.0546 (\mathcal{M}/M_{\odot})^{-5/3} (\nu(f)/{\rm yr^{-1}})~(f/\rm Hz)^{-11/3}~~~ \rm Hz^{-1}.
\end{split}
\label{eq_dndf}
\end{equation}
Hence the number distribution has contributions from the birth-rate  $\nu(f)$, which
is important when significant numbers of young DDs are being injected into the population at high frequencies,
and from the orbital decay lifetime which gives the $f^{-11/3}$ distribution and which is important when short-period
DDs have all evolved from old long-period DDs.

To illustrate, we assume that $\nu(f)$ and $\mathcal{M}$ are independent of frequency $f$.
If $\mathcal{M}=0.5$ $M_{\odot}$ and $\nu=0.004$ yr$^{-1}$ (average values
estimated from simulations with quasi-exponential SF in the frequency range
$10^{-5}-10^{-1.6}$ Hz), the number probability of DDs  $\frac{{\rm d} n(f)}{{\rm d} f}\approx0.0007f^{-11/3}$
(${\rm d} f=1$ yr$^{-1}$). This relation is plotted in Fig.\,\ref{fig_logh}
and has the same slope as the simulation with instantaneous SF (approximately a $\delta$-function)
for $\log f> -2.8$.

The differences between the number distributions of resolved DDs in the three SF models can be
understood simply in terms of the SF history.
The DD birth rate $\nu(f,t)$ can be expressed as
$\nu(f,t_{\rm disc}) =\int_{0}^{t_{\rm disc}} \ddot{C}_{\ast}(f,t_{\rm disc}-t_{\rm sf})S(t_{\rm sf}) {\rm d}t_{\rm sf}$
where $t_{\rm disc}$ is the age of the thin disc, $t_{\rm sf}$ is the time of star formation,
$\ddot{C}_{\ast}$ is the contribution function, and $S$ is the star formation rate.
$\ddot{C}_{\ast}$ provides information about the number and frequency distribution of DDs
from a given epoch, and is independent of the SF history (see \S\,2.1 and Figs. 6-7 in
\citet{Yu11} for details). The DD birth rate is obtained by convolution with $S$.

If $S$ is a $\delta$-function, the birth rate of DDs after 10\,Gyr is a very weak function of $f$.
Most present-day close DDs would be long-evolved DD systems,
so their number distribution follows the power-law relation $f^{-11/3}$ (Fig.\,\ref{fig_logh} ).
However, if $S$ is continuous over 10\,Gyr, contributions at every SF epoch must be included and
will increase the number of DDs at high frequency.

Thus, assuming that (a) all other inputs to the BSPS model are valid (or at least that systematic
errors can be eliminated by other means)
and (b)  there is no significant contamination by resolved sources from other locations,
it is evident that GW observations of DDs could distinguish between different
SF histories. In particular, the frequency above which the distribution develops
 a $f^{-11/3}$  gradient is indicative of the time since the last major SF epoch.

\section{Discussion}
\label{sec_discussion}

The impact of different SF histories provides one reason for the differences
between calculations of DD number distributions and GW strain amplitude-frequency relation
by different authors \citep{Evans87,Nelemans01b,Ruiter09,Yu10}, especially where these authors
adopt similar approximations for the physics of binary star evolution.
The method of interpolation (see \S\,\ref{sec_simulation}) is important for predicting the absolute number
of resolved DDs but not crucial for the relative numbers in different models.
In the present simulation, we only interpolate to produce a complete Galactic sample from
the more limited BSPS sample. There is no extrapolation and the interpolation errors
decrease the predicted GW strain and the number of resolved DDs by no more than
$\sim$30\%.

\citet{Nelemans04} estimate that there may be $\approx2.13\times10^8$ close detached DDs in the
Galactic bulge and disc,  with $\approx2.3\times10^7$ semi-detached DDs, assuming
an age of 13.5 Gyr and a SF history following \citet{Boissier99}.
We find $\approx2.4\pm0.5\times10^8$  DDs in the bulge and thin disc,  with
$\approx5\times10^5$ semi-detached DDs ($\approx0.3\%$ of all DDs in the thin disc) \citep{Yu10,Yu11}. The average birth rate of DDs given by \citet{Nelemans04}
($1.58\times10^{-2}$\,yr$^{-1}$) is slightly less than our result ($2.4\times10^{-2}$\,yr$^{-1}$).
However, we give a much lower average birth rate for semi-detached
DDs $\approx5\times10^{-5}$\,yr$^{-1}$, compared with  $\approx1.7\times10^{-3}$\,yr$^{-1}$ by
\citet{Nelemans04}. The local space density of semi-detached DDs in our quasi-exponential SF model
of the thin disc is $\approx6.2\times10^{-7}$ pc$^{-3}$, slightly less than the constraint of
 $1-3\times10^{-6}$ pc$^{-3}$
obtained from the Sloan Digital Sky Survey \citep{Roelofs07} on the assumption that
semi-detached He+He and CO+He DDs look like AM\, Canum Venaticorum binaries.

\citet{Ruiter10} estimate there to be $\approx2.48\times10^{7}$ DDs with orbital
periods in the range 200 s to 5.6 hr in the Galactic disc,
including $\approx1.62\times10^{7}$ semi-detached DDs ($\sim$65\% of all DDs).
They assume a constant SF rate (4 $M_{\odot}~{\rm yr}^{-1}$) in the disc for 10 Gyr.
The average birth rate of DDs in this period range is $\sim2.48\times10^{-3}$\,yr$^{-1}$,
while the average birth rate of semi-detached DDs is $\sim1.62\times10^{-3}$\,yr$^{-1}$.
The latter is significantly higher than our result. The local density of semi-detached DDs
in \citet{Ruiter10}, $2.3\times10^{-5}$ pc$^{-3}$, is substantially higher than both our value
and the observation. The formulae adopted to simulate Galactic structure may contribute
a significant fraction of the differences in estimates of the local density of DDs; \citet{Nelemans04},
 \citet{Ruiter10}, and ourselves adopted density distributions $\rho\varpropto {\rm sech}(z/h_{z})^{2}$,
$\rho\varpropto e^{(\mid-z\mid/h_{z})}$, and $\rho\varpropto {\rm sech}^{2}(-z/h_{z})$ respectively,
where $\rho$ is  mass density, $z$ the height of the thin disc in the cylindrical coordinates, and $h_{z}$
the scale height of the thin disc \citep{Yu10}.

\citet{Nissanke12} recently confirmed that thousands of detached DDs may be detected by
space-borne GW detectors with 5 and 1 Mkm arms, but there may be only a few ten to a few hundred
 detections of interacting DD systems. This is attributed to an assessment of detection prospects,
based on iterative identification and subtraction of bright sources with respect to both instrument
and confusion noise. Our results also imply that detached DDs will dominate the GW
 signal from the total Galactic DD population in the GW frequency range of $10^{-5}-10^{-1.6}$\,Hz.
 However, they also argue that the SF history of the  disc is an important factor and
may affect the number ratio of the detached and semi-detached DDs detected by a GW observatory.

Since the GW background may be affected by sources including DDs beyond the Milky Way, an estimate of contamination from the Local Group would be of interest. Observations indicate that SF history in the Local Group is complicated. For instance, the Large Magellanic Cloud (LMC) most likely experienced a dramatic decrease of SF following an initial star burst some 12 Gyr ago. SF resumed some
5 Gyr ago  has since proceeded with a low average rate of roughly $0.2~M_{\odot}{\rm yr}^{-1}$,
 peaking roughly 2 Gyr, 0.5 Gyr, 0.1 Gyr and 0.012 Gyr ago \citep{Harris09}.
The Small Magellanic Cloud (SMC) has a similar history to the LMC, but with a different look-back
time for the initial star burst ($\approx8.4$ Gyr) and with minor star bursts roughly 2.5, 0.4, and
0.06 Gyr ago \citep{Harris04}. Our results (e.g. Fig.\,\ref{fig_sfgws}) show that recent (2 Gyr) SF has a
 significant contribution to the GW signal from DDs in the Galactic thin disc. Assuming that DD
 contribution functions are the same for both the Galaxy and for the LMC and SMC, the GW signal from the
thin-disc DDs may well be contaminated by DDs in other Local Group galaxies. This should
be explored in the future.

\section{Conclusion}
\label{sec_conclusion}

Monte Carlo simulations have been carried out to synthesize
the GW signal from DDs based on a binary-evolution model
\citep{Han98,Hurley02,Yu10}, different star-formation models and
the present thin disc structure of the Galaxy.
The GW signal consists of a continuum at low frequency and a
discrete signal at high frequency. The continuum arises mainly
from DDs with main-sequence progenitors formed in the early disc
(i.e. between 0 and 8 Gyr). Recent SF (i.e. between 8 and 10 Gyr) gives
rise to a significant fraction of short-period DDs corresponding
to the GW signal at high frequency, which is mostly resolved and
which is thus sensitive to the recent star-formation history.
At the highest frequencies, where the orbital-decay
timescale governs the evolution, we have shown that our Monte Carlo simulations
reproduce the distribution expected from analytic approximation.
Both the intercept of the high-frequency distribution
of the resolved DD population and  the frequency below  which
the distribution drops below that predicted by the orbital-decay
timescale are sensitive to recent star formation.

Since the high-frequency GW signal is dominated by DDs from recent star formation,
future work should explore whether GW observations can be used to constrain the
time since the most recent major burst of star formation in the Galaxy.

\section{acknowledgements}
The Armagh Observatory is supported by a grant from the Northern
Ireland Dept. of Culture Arts and Leisure. We thank the referee for the constructive suggestions and comments.

\bibliographystyle{mn2e}
\bibliography{ddgwr}

\label{lastpage}
\end{document}